# A First Step in the Co-Evolution of Blockchain and Ontologies: Towards Engineering an Ontology of Governance at the Blockchain Protocol Level

Henry M. Kim
Schulich School of Business, York University, 4700 Keele St., Toronto, Ontario Canada
hmkim@yorku.ca | (416) 736-2100

Marek Laskowski
Schulich School of Business, York University, 4700 Keele St., Toronto, Ontario Canada
marlas@yorku.ca | (416) 736-2100

Ning Nan
Sauder School of Business, University of British Columbia, 2053 Main Mall, Vancouver, British Columbia, Canada
ning.nan@sauder.ubc.ca | (604) 822-8500

## Abstract

At the beginning of 2018, there is a growing belief that blockchain technologies constitute a revolutionary innovation in how we transfer value electronically. In that vein, blockchain may be a suitable complement to ontologies to achieve a big part of Tim Berners-Lee's vision of the semantic Web. We believe that if this complementarity is to be achieved blockchain and ontologies must co-evolve. In this paper, we focus on what and how to engineer models, methods, designs, and implementations for this co-evolution. As a first step in this co-evolution, we propose a conceptual design of a governance ontology represented as meta-data tags to be embedded and instantiated in a smart contract at the blockchain protocol level. We develop this design by examining and analyzing smart contracts from the infamous The DAO experiment on the Ethereum blockchain. We believe there are two contributions of this paper: it serves to inform and implore the blockchain and ontology communities to recognize and collaborate with each other; and it outlines a roadmap for engineering artifacts to bridge the gap between blockchain community's focus on protocol-level blockchain interoperability and the ontology community's focus on semantic-level interoperability.

Keywords: blockchain, ontologies, smart contracts, blockchain interoperability, governance, Ethereum

# A First Step in the Co-Evolution of Blockchain and Ontologies: Towards Engineering an Ontology of Governance at the Blockchain Protocol Level

## 1. Introduction

Awareness of blockchain technologies has been driven by the exponential growth in valuations of Bitcoin, Ethereum, and other cryptocurrencies, and frequent announcements of yet another industrial application of blockchain. Yet, the core premise of blockchain is straightforward: Rather than relying upon a trusted intermediary to proprietarily maintain one ledger of transactions between members of a network, allow all members to maintain their own copies of the same ledger and ensure that the copies are all synchronized. This would obviate the need for the intermediary, who often exploits its information asymmetry advantage to act in self-interested and extractive, inefficient, or corrupt ways. Bitcoin represents the best example of blockchain in action and a great example of how blockchain disrupts the intermediary is the following: In 2016, a bitcoin-based remittance system could transfer $100 from Canada to Philippines nearly immediately for one dollar. In contrast, a traditional bank charged a ten dollar fee and took three days to clear the payment.

A complementary benefit of blockchain is that it can also be used in cases where no natural intermediary exists. A large-scale application of blockchain for traceability represents an example. A global food safety blockchain consortium involving IBM, Walmart, Unilever, Nestlé, Dole, Kroger, JD.com, and others was formed at the end of 2017 to collaborate on a variety of different projects (del Castillo 2017a)(Higgins 2017a). Unlike financial networks, the presence of intermediaries or hubs that maintain ledger of transactions in supply chain networks is not the norm. Therefore, especially for complex global supply networks, intermediaries responsible for tracking goods such as food, pharmaceuticals, or consumer goods where tamper-proofing or demonstrating provenance is critical have never existed in any large scale.

Before the pseudonymously named Satoshi Nakamoto launched Bitcoin and blockchain with their whitepaper (Nakamoto 2008), there were already thriving communities developing decentralized systems that could inter-operate in the absence of an intermediary—the ontology and open data communities. The development premise for these communities assumes that data from disparate sources is "open" to being shared, and that semantics in the form of ontology representations and meta-data are required to unambiguously interpret data that is processed for inter-organizational business processes and Web services. The premise for Bitcoin and traceabilty blockchains assumes that blockchain is needed because data is closed for sharing; that is, it is assumed that self-interested intermediaries tend to hoard and close-off data about network members' transactions or complex business networks like supply chains tend to have no natural intermediaries and hence there is no network-wide record of transactions.

In most enterprise contexts, the assumption that data is closed for sharing is the default. Google, Facebooks, banks, and others own their data and choose to make some data available to others at their discretion via API's and Web services. They constrain their data with their own meta-data so the need for others to define semantics is not compelling. Understandably, they do not expose so much of their data that they lose their information asymmetry and scale of data advantages. So, data sharing is done in very limited way without the need for ontologies within contexts where self-interested intermediaries exist. On the other hand, many government agencies are proponents of open data and do make their data available. Others then define semantics for this data [e.g. (Wang and Fox 2016)(Golbeck et al. 2011)]. However, commitment to maintaining and updating open data varies widely across different agencies. Hence data sharing using ontologies may also be limited within contexts where intermediaries that make their data available are ineffective or inefficient, such as some government agencies. Finally, ontologies ostensibly ought to be useful when there are no data intermediaries, such as in supply chain networks. Even though there are widely adopted standards like GS1 for retail barcodes, and industry-specific United Nations CE/FACT standards for e-business, at the semantic level beyond the data standards level, there are arguably no widely adopted supply chain ontologies.

We have outlined how practical ontology use is limited in contexts where the data upon which ontology semantics is to be applied is controlled by self-interested, extractive, ineffective, or inefficient intermediaries, or in contexts where intermediaries do not readily exist. These are the same contexts for which blockchain use is especially touted. Therefore, we posit that blockchain adoption will lead to greater ontology adoption; we posit that they will co-evolve.

In the next section, we will present elementary blockchain concepts and explore our statement further. In Section 3, we will discuss one specific implication—the need for an ontology to describe the governance structure of blockchain smart contracts. In Section 4, we present a proof-of-concept design, to ultimately be executed on the Ethereum application development platform and the *Inter-Planetary File System* (*IPFS*), and encoded in the *Solidity* language. In Section 5, we make concluding remarks and discuss future.

## 2. Blockchain and Ontologies

### 2.1. Blockchain and Smart Contracts

Blockchain is a revolutionary technology, which leverages the same cryptographic principles that underlie secure computer communications to implement a shared ledger or database that delivers a "single version of the truth" amongst numerous, sometimes adversarial, stakeholders. Data and business processes that use the Blockchain gain a considerable level of auditability, and security. Blockchain itself enables value to be transferred over the Internet, while smart contracts which are programs that encode business logic that are executed by the Blockchain can be used to implement autonomous agreements between stakeholders.

Blockchain is a peer-to-peer public ledger maintained by a distributed network of computational nodes. The most common design as used by far the two largest implementations—Bitcoin and Ethereum networks—works as follows. A peer's request to perform a transaction with another peer is broadcast to blockchain nodes. A collection of such transactions within a period of time is verified by an algorithmically-chosen node, called a miner, who is not a party to any of these transactions. The miner verifies ("mines") that the transactions are valid; for instance, for Bitcoin, miners ensure that a payer cannot double spend and overdraw from their account. This collection of verified transactions is recorded notionally as a block of data. A subsequent block records transactions verified by another algorithmically-chosen miner within the next time period. For one block, a hash value is calculated as a function of the block's transactions data as well as the hash value of the previous block. This embedding of hash values serves to daisy chain the blocks. This design also makes tampering with recorded blocks nearly impossible.

If records in a historical $n^{th}$ block are falsified, then the new calculated hash value for that block will be different from the recorded value. Part of the verification of the current block [an $(n+i)^{th}$ block] requires the miner to check for such incongruence in previous blocks, so the tampering attempt will be quickly identified. In fact, to not get caught, the perpetrator must meticulously falsify $1^{st}$, $2^{nd}$, and $3^{rd}$, to $(n-1)^{th}$ and $n^{th}$, to $(n+i-1)^{th}$ block before the $(n+i)^{th}$ block is verified. At the end of 2017, Bitcoin is mined every ten minutes and the network is at around its $520{,}000^{th}$ block; Ethereum mining occurs on average every 15-20 seconds and the network is at around its $480{,}000^{th}$ block. Given these figures, tampering is reasonably intractable.

A smart contract is merely a computer program that executes in a blockchain environment. The term "smart contracts" is a misnomer. There is nothing inherently "smart" nor contract-like about a blockchain smart contract, though a well-written smart contract can autonomously enforce a contractual obligation—e.g. automatically sending payment from a payer once the recording of a service provided by the payee as per the logic programmed in a smart contract is verified on the blockchain. Smart contracts are functionally expressive: For instance, the Solidity programming language widely used to code smart contracts in Ethereum is Turing complete. There are however differences compared to more fully-expressive languages like Java and Python. Certain paradigms like dynamic typing and non-deterministic execution are not permitted in smart contracts as they are inconsistent to cryptographically secured and tamper proof design principles of blockchain.

## 2.2. Blockchain and Ontologies: Co-Evolution

There have not been that many works that examine blockchain and ontologies. One area of investigation has been in developing a characterization of blockchain concepts and technologies as ontologies. These works present an ontology of blockchain as a natural language characterization of blockchain constructs—e.g. "What are the different kinds of blockchains?" (Glaser 2017)(Tasca et al. 2017). Or, commit further to representing an ontology of blockchain as conceptual modelling formalism—e.g. representing that "an EconomicAgent initiates a Transaction, which is governed by a SmartContract" in UML notation (de Kruijff and Weigand 2017a)(de Kruijff and Weigand 2017b).

Rather than developing an ontology of blockchain, other works endeavor to represent ontologies of enterprise or general concepts within a blockchain context, and furthermore, commit to prototypical implementations. A comprehensive treatise on integrating blockchain and ontologies is the BLockchain ONtology with Dynamic Extensibility (BLONDiE) project (Ugarte 2017).

The BLONDIE ontology implements an ontology in OWL that interfaces with Bitcoin and Ethereum networks. The implementation answers competency questions such as "Who was the miner for each block?" and "How many transactions were included in each block?" The project also presents a Decentralized Supply Chain Application (DeSCA), which is a prototype where different companies along a supply chain would use a common blockchain as source of data and inter-organizational business processes would entail use of ontologies represented using an RDF framework. Another comprehensive effort comes from the Blockchain.Lab.Yorku.ca. They propose and develop a prototypical implementation of an ontology of traceability as an Ethereum smart contract for a supply chain provenance use case (Kim and Laskowski 2016).

Beyond implementations, these two projects also speculate about the co-evolution of blockchains and ontologies. Ugarte shares Kim and Laskowski's vision: "Semantic Web is an incomplete vision so far; but a homogeneous revolutionary platform as a network of Blockchains could be the solution" (Ugarte 2017, p. 1). We believe the first step towards this vision is the recognition that blockchain and ontologies ought to co-evolve. That is, as we have already outlined, recognition by the semantic Web community that willingness for different organizations to share data is greatly enhanced when it is cryptographically secured and its truth is verified by a blockchain is needed. Works presented so far ostensibly recognize this. Recognition by the blockchain community that sharing data requires ontologies is also needed. The recognition that sharing data across different blockchains is very important has clearly registered with the blockchain community, even if mention of ontologies has not. In May 2017, leading representatives from the four largest blockchain platforms—Bitcoin, Ethereum, Hyperledger, and R/3's Corda—said: "All of these individuals on stage with us are all building technologies with which we would someday like to interoperate" (del Castillo 2017b).

Indeed, there are many efforts at enabling inter-chain interoperability (Ethereum's AI 2017) and two of the most developed are Cosmos (Kwon and Buchman 2017) and Aion networks. Cosmos is conceptualized as an Internet-like network of public blockchains organized in hubs and zones, where each zone maintains its own blockchain and "plug" into hubs. Aion network (Higgins 2017b)(Spoke and Nuco Engineering Team 2017) connect both public and private blockchains, but rather than a hub-and-zone architecture, it acts as a common translation blockchain that inter-operates between heterogeneous blockchains. There are many differences between the two, but what they have in common is their emphasis on protocol-level details of blockchain interoperability, such as mechanisms for achieving consensus that data transported from one blockchain to another has been verified in the destination blockchain, and tokenization of an inter-blockchain cryptocurrency system used to incentivize third party verification. What is hardly addressed though is semantic interoperability. This means that in the approach of Aion and Cosmos a piece of data can be verified and transported from one blockchain to another—and a third party may receive cryptocurrency payment for verification—but without the semantics of its proper use transported as well.

We believe that between protocol-level inter-operability designed with focus on issues like consensus and tokenization mechanisms and semantic level inter-operability with focus on knowledge engineering and representational language there is a protocol/semantic interface. Ugarte insightfully recognized the need for this layer. He stated that there are three possible approaches to "semantify" the blockchain: 1) Encode RDF class definitions within the blockchain, or maintain the definition elsewhere "off-chain" but place hashes of those definitions only on the blockchain; 2) Extend data interchange protocols of an interoperating network like Aion or Cosmos to represent RDF constructs; or, 3) Map blockchain data instances directly onto appropriate RDF classes that are not on the blockchain (i.e. off-chain).

It appears that Ugarte uses the last approach in his DeSCA prototype. In the context of Aion and Cosmos's protocol level inter-operability focus, the difficulty of Ugarte's approach is that this mapping would not necessarily be maintained by Aion or Cosmos when they transport data between blockchains. In order to preserve these mappings, Aion or Cosmos would need to be able to access and process RDF encoded ontologies off-chain, and not only do they not facilitate this now but are unlikely to do so because their focus is primarily at the blockchain protocol level. What we are proposing with our work is an approach that is more complementary to their efforts, and which entails a hybrid of approaches 1) and 2). We propose to: encode ontology constructs within the blockchain at the protocol level as specially tagged hashes that point to semantics (ontology definitions and axioms) kept off-chain.

Figure 1: Schematic of Different Approaches to Representing Semantics for Blockchain

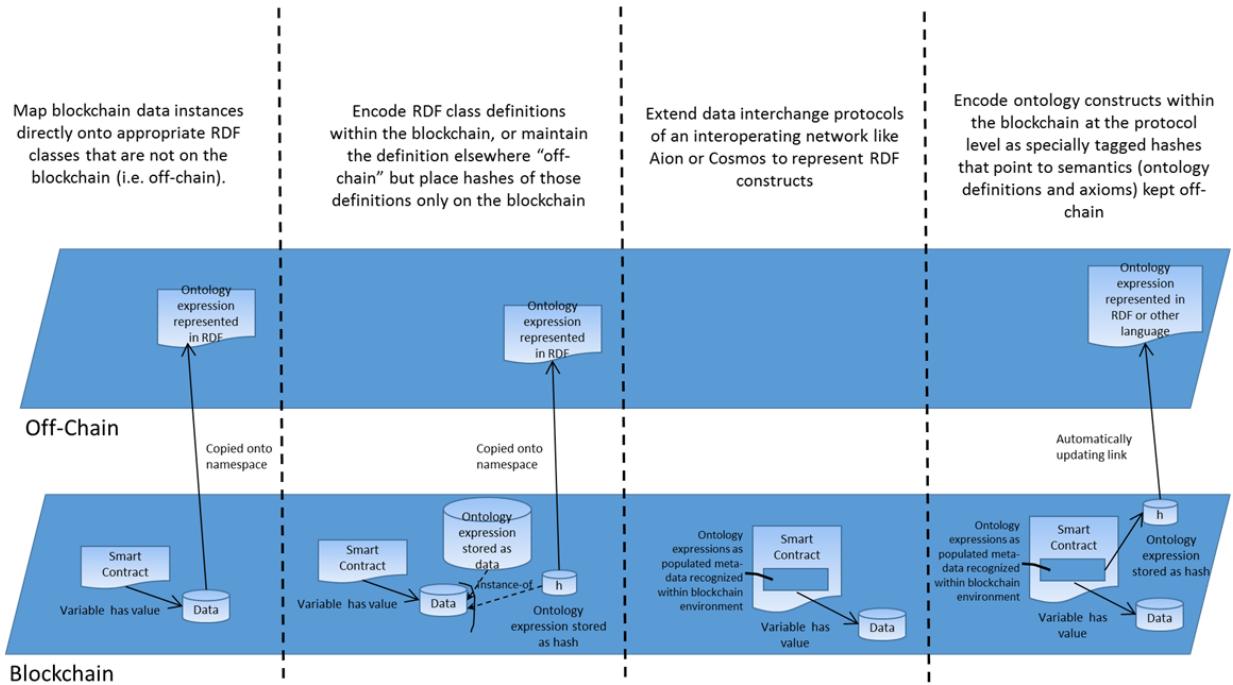

## 3. Ontologies for Governance on the Blockchain

Encoding ontologies on the blockchain entails ontological commitment. That is, there would be an expectation that blockchain developers adopt some ontology terms as meta-data. For users of a public, domain-independent blockchain like Ethereum, this is a large commitment. Therefore, our inter-operability approach cannot be used for any arbitrary domain ontology. Only those domains that are very core to the function of the blockchain ought to be considered. We posit that the domain of organizational governance is such a core domain. With respect to the Ethereum network, for instance, our approach would entail motivating smart contract writers to use special governance ontology meta-data tags within their Solidity source code.

Why would blockchain programmers volunteer to adopt an ontology of governance? Because, governance—or more appropriately autonomous organizations capable of self-governance—is a concept inexorably tied to the early days of the Ethereum blockchain. Specifically, The DAO (Decentralized Autonomous Organization) was a spectacular failed experiment in autonomous self-governance. Because of this tie, if there were any meta-data tags that programmers would be willing to use in their smart contracts, it would be ones that are part of an ontology of governance.

The DAO was meant to function as an autonomous venture capital fund where investors invested capital through the DAO and then voted to allocate their capital to different Ethereum-based projects that were presented to them (DuPont 2018). Votes were to be tabulated, projects were to receive funding after receiving sufficient funding, and account balances were to updated using pre-defined rules expressed and executed using a smart contract. However, shortly after its initiation in the summer of 2016, The DAO was hacked and 3.6M Ethers was cordoned off by a hacker, inaccessible to their rightful owners. Ultimately, Ethereum users voted to "hard fork," and return the state of Ethereum just prior to the hack. All funds were returned, and The DAO, shuttered (Mehar et al. 2017). Around the time of the hack, the stolen Ethers amounted to between $50M - $60M USD. As per valuations as of January 2018, the stolen Ethers would be worth around $3.6B USD!

It is unfortunate that The DAO experiment never became an on-going concern. We could speculate that the smart contracts representing those projects that would have sought funding from DAO investors would evolve to incorporate constructs that are akin to our governance ontology. What is the basis for that speculation? The DAO designers specified mechanisms for voting to invest in prospective projects but did not prescribe to how investors would evaluate the projects. Literature states that investors evaluate venture investing according to dimensions such as innovative technology, potential for rapid growth, a well-developed business model, and an impressive management team (Ross et al. 2016). Of these dimensions, management team is clearly the most straightforward concept to represent within a smart contract. That is, it is reasonable that clever entrepreneurs seeking to receive funding from The DAO investors would embed structured descriptions of their organizational structure including their management team within their smart contracts, similar to how IPO prospectuses prominently feature a company's governance structure. Even though The DAO died, there are DAO-like endeavours that are cropping up. Outside of use cases for venturing or crowd funding, there are blockchain use cases that entail autonomous management of IT infrastructure or autonomously coordinating operations between supplier chain partners, all via smart contracts (Dhillon et al. 2017).

To recap, we have posited that a pragmatic first step in co-evolving blockchain with ontologies is to encode ontology representations as meta-data tags to be embedded and instantiated in a smart contract at the blockchain protocol level. We further posit that an ontology representing governance structure is more likely to be adopted by smart contract writers than ontologies representing other domains.

There are severe restrictions on representing an ontology on the blockchain. The programming language used for blockchain has some restrictions on expressions and is procedural. Also, blockchain computations occur only between one verified system state to the next verified system state. Such stepwise procedural flow is inconsistent with inferences on declarative expressions. This means that a lot of inference capabilities normally associated with ontology use cannot naturally be done in a blockchain environment. Nevertheless, we present the TOVE Organization Ontology as our inspiration. For our future work, we aim to simplify and tailor it to fit our approach, though admittedly it is little used in this prototypical implementation.

Figure 2: TOVE Object-Oriented Model of Organizational Structure

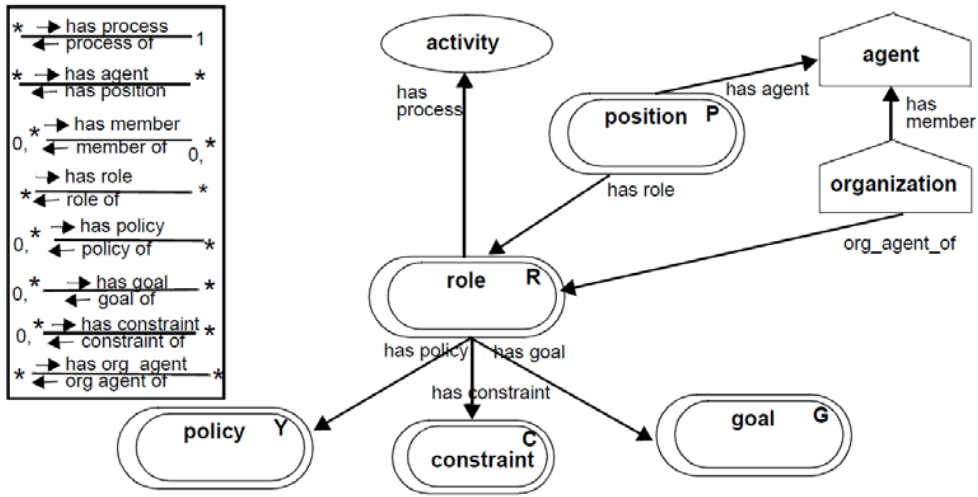

Figure 3: TOVE Object-Oriented Model of Information Flows

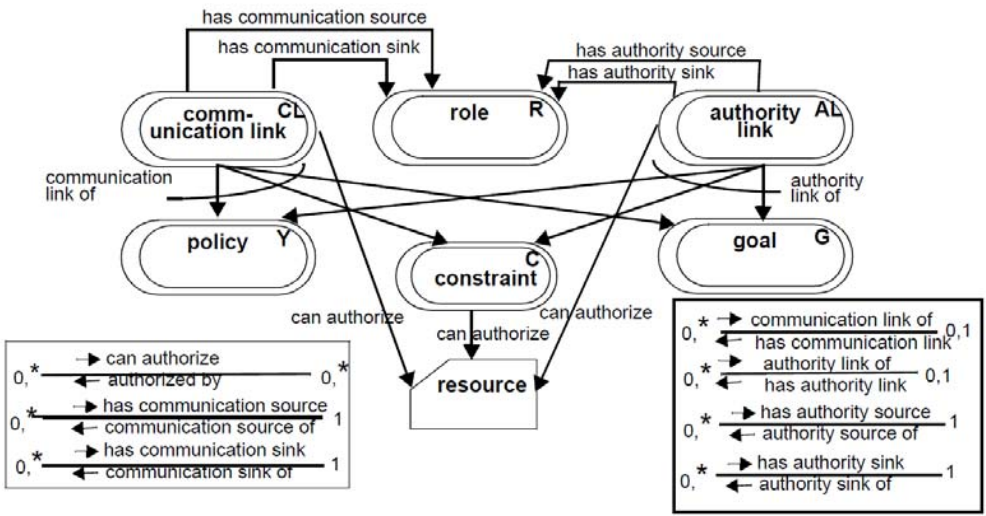

## 4. Prototype Design

Next, we initiate encoding ontology representations as meta-data tags to be embedded and instantiated in a smart contract at the blockchain protocol level through a case study of the actual The DAO contract.

Figure 4: Role/Goal Model of The DAO smart contract governance

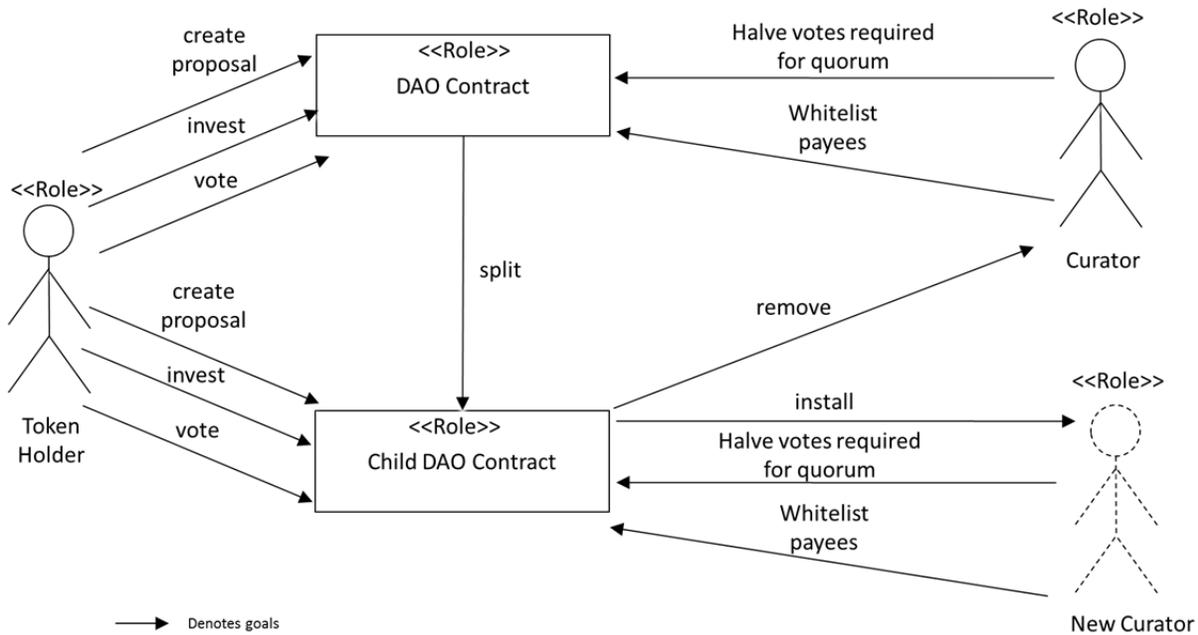

The governance structure of The DAO contract (https://etherscan.io/address/0xbb9bc244d798123fde783fcc1c72d3bb8c189413#code) described in Figure 4 is both complex and relatively novel. Unlike many smart contracts, The DAO does not have the role of an owner (typically the entity that paid to deploy the smart contract) that has privileged control over the behaviour of the contract. Instead, there is a curator appointed when The DAO smart contract is created. The curator manages a whitelist of addresses that The DAO is permitted to send ether to, acting as a sort of gatekeeper to whitelist viable proposals. The curator can be an individual or could itself represent another smart contract with its own internal governance structure.

Besides the curator, the main type of entity that interacts with the smart contract is the Token Holder. Individuals become Token Holders by sending Ethers into The DAO contract, which subsequently grants them rights to create new proposals and vote on proposals. Also, it's interesting to note that the curator has a second power: to reduce the number of Token Holders required to establish voting quorum, presumably to avoid deadlock.

Neither the creator of The DAO smart contract nor the Token Holders have the ability to change the curator of the original DAO. However, Token Holders can propose to split the DAO into the original and a new "child" DAO to which a new curator can be assigned. This serves as a "check and balance" to the power of the curator; if there are enough unhappy Token Holders they have the ability to split off and appoint a new curator over their portion of the funds.

Had The DAO not collapsed, the curator may have been key in mitigating potential social or economic malicious activity, via their role as gatekeeper. It should be noted that curators could not have prevented the attack on The DAO since the curator is not in charge of the withdrawal function in which the bug was used to exploit The DAO.

In addition to the governance structure of The DAO, other governance structures for Smart Contracts are emerging. To address the complexity of an evolving ecosystem of governance structures, we propose a protocol which includes machine parseable representation of smart contract governance. The protocol ensures that the most up to date and accurate description of a particular smart contract's governance structure is verifiable through the smart contract instance itself. This in turn enables extensive off-chain reasoning through annotated smart contract code, as well as limited on-chain reasoning performed by a first smart contract concerning governance structure of a second smart contract.

To this end, we describe the protocol in the context of the Ethereum ecosystem, as it represents one of the largest public smart contract blockchains by transaction volume, valuation, and trading volume. However, these principles could equally well be applied to other blockchain implementations and ecosystems.

Figure 5: A Protocol to infer Smart Contract governance structure

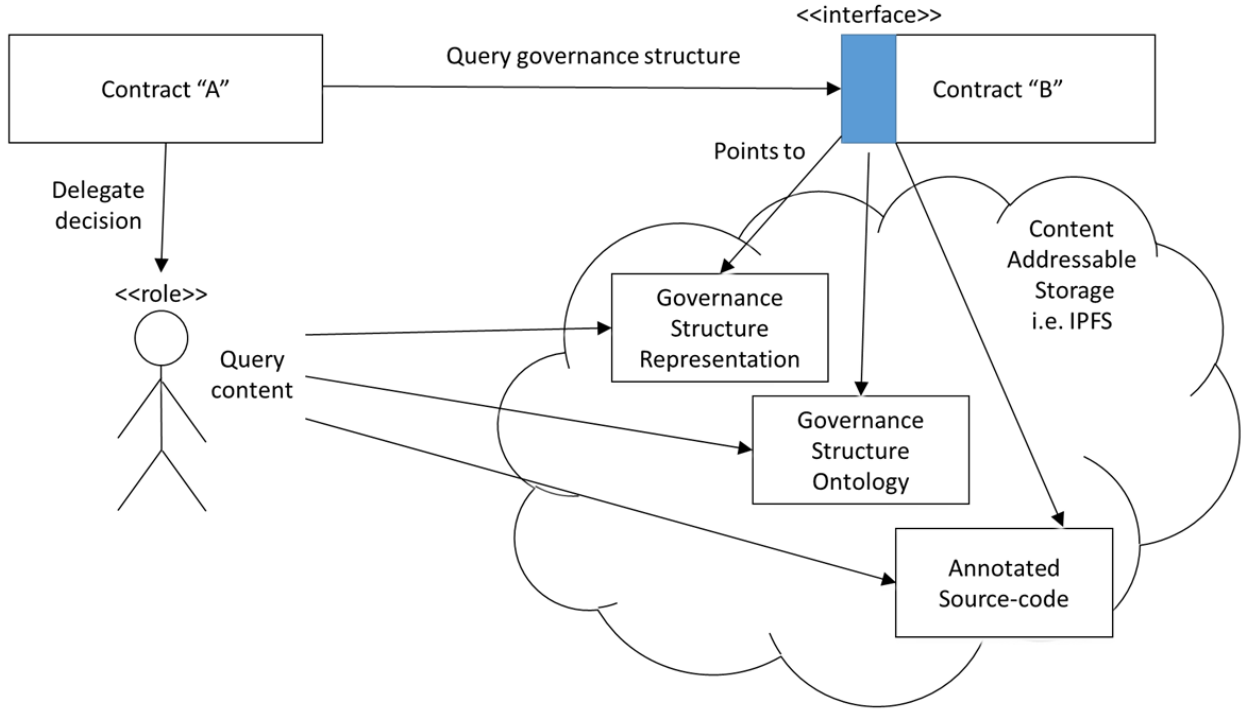

With regards to Figure 5, we assume that Contract A needs to reason about the governance structure of a second Contract "B" to assess, for example, counter-party risk. To enable this use case, an interface is provided by B to permit A to query a "fingerprint" (formally known as a hash or hash-digest) of the full governance documentation. The choice to use hashes here is a pragmatic one, in response to the relatively resource constrained environments of contemporary blockchains. Hashing functionality is already built into Ethereum as a cryptographic primitive, and suffices for efficiently determining whether two complex documents are identical while limiting the amount of data that needs to be put into the blockchain.

In some cases, Contract A may be able to perform all the requisite inference locally, on-chain, deciding how to proceed based on its internal business logic. This will be the case when both A and B are referring to of the same set well of known governance templates and share the same representational ontologies. The "fingerprint" of the organizational structure returned by the query is enough to determine whether the governance structure is known to Contract A. If so, the logic can be a simple if-else construct based on the declared governance structure. Other times it may be undecidable by the internal logic of the Smart Contract due to a novel governance structure that has not yet been considered, or an irreconcilable difference in underlying ontological representation. In these cases Contract A can delegate this decision making off-chain to an external agent which could be a human or artificial decision maker. In this case, more detailed information can be retrieved and processed by an off-chain Agent known to Contract A. This data is made securely accessible through a "Content Addressable Storage" network (i.e. IPFS) using the same hashes returned through Contract B's interface.

It is expected that contract creators will voluntarily and truthfully disclose the governance structure in order to provide transparency to potential customers. Contracts that are discovered to not correctly implement their claimed governance structure could be viewed as either buggy or malicious. In any case, such smart contracts (and authors) may lose reputation and eventually be "blacklisted" by the ecosystem.

A basic definition of the interface referred to in Figure 6 is contained in Code Listing 1 in the Appendix. An example of another contract querying that interface in order to perform some basic inference is demonstrated in Appendix: Code Listing 2. It's relatively straightforward to imagine more elaborate scenarios where the contract performing the inference can take a different branch of execution depending on the governance structure returned by the other contract. Finally, an example of entity annotations in Solidity code is shown in Appendix: Code Listing 3, based on the widely used Doxygen style. In the style of this convention, entities in the Solidity Smart

Contract that are related to the governance of the Smart Contract are annotated with a new "@OntoInstance" tag that extends the common tag types found in Doxygen annotations.

## 5. Concluding Remarks

At the beginning of 2018, there is a growing belief that blockchain technologies constitute a revolutionary innovation in how we transfer value electronically. In that vein, blockchain may be a suitable complement to ontologies to truly achieve Tim Berners-Lee's vision of the semantic Web. We believe that if this complementarity is to be achieved blockchain and ontologies must co-evolve. In this paper, we focus on what and how to engineer artifacts (models, methods, designs, and implementations) for this co-evolution.

Specifically, we posit that a pragmatic first step in co-evolving blockchain with ontologies is to encode ontology representations as meta-data tags to be embedded and instantiated in a smart contract at the blockchain protocol level. We further posit that an ontology representing governance structure is more likely to be adopted by smart contract writers than ontologies representing other domains. Finally, we analyze the smart contracts of the infamous The DAO experiment on Ethereum, from which we outline a conceptual design for a governance ontology at the blockchain protocol level.

There is much more work to be done. As a start, the prototype design outlined in this paper must be implemented in Ethereum or another blockchain environment. Also, a more complete analysis of smart contract governance structures, including development of a more structured ontology remains to be done. This paper serves as an initial roadmap for engineering artifacts for blockchain/ontology co-evolution, and there are many design issues that need to be more fully explored. One clear area of exploration is how ontology constructs embedded at the protocol level on the blockchain and more formal ontology representations at an RDF/OWL/SPRAQL semantic level would interface in a live, scalable interface.

Future work notwithstanding, we believe there are two contributions of our work. One, our statement that blockchain and ontologies ought to co-evolve serves to inform both the blockchain and ontology communities and implores them to recognize that their endeavours are complementary and would benefit from greater collaboration. Two, we outline a roadmap for engineering models, methods, and designs to bridge the gap between blockchain community's focus on protocol-level blockchain inter-operability and the ontology community's focus on semantic-level interoperability.

# Appendix

## Code Listing 1

```
/*solidity-like pseudocode*/
contract Inference {
    //SHA256
    bytes32 standardDaoHash =
0x56B3CAEC460654DF16232C6207DFF138807401A2D4
39366BFEB80E981C1410BF;
    bytes32 standardBlockchainOntologyHash =
0x329A16326F27E5C96A45BC1DDA19C49C18ED32593E
868F7B2534CDAE3A0A0665;
    bytes32 annotatedSourceCodeHash =
0x0F9E141DA2589A9EE28BC3F825E03D08F8F69F5AA9
530ED485E5A19904C52579;

    function getGovernanceModel() public
returns (bytes32)
    {
        return standardDaoHash;
    }

    function getReferenceOntology() public
returns (bytes32)
    {
        return standardBlockchainOntology;
    }

    function getAnnotatedSource() public
returns (bytes32)
    {
        return standardDaoHash;
    }
}
```

## Code Listing 2

```
/*solidity-like pseudocode*/
function isFamiliarGovernance(address
otherContract) returns (bool)
{
    //in this example we're only familiar
with one kind of governance model
    //and one ontology
    //however, we can imagine a more complex
data structure here....
    Inference otherInference =
Inference(otherContract);
    if(otherInference.getGovernanceModel()
== standardDaoHash
    && otherInference.getReferenceOntology()
== standardBlockchainOntology){
        return true;
    }
```

```
    return false;
}
//we could imagine complex business logic to
behave differently, depending on the
governance structure of the other
//if not a match, could pause, and wait for
off-chain intervention
```

## Code Listing 3

```
/**                          @ontoInstance
0x5C15E9701E5B866B92C31EE4CB0CDD767024A9091D
B39045310E1FB376DB1A65
*/
address curator;
//Doxygen style annotation of "curator"
address within the smart contract
```